# Intrinsic electron mobility exceeding $10^3$ cm$^2$/Vs in multilayer InSe FETs

*Sukrit Sucharitakul [†], Nicholas J. Goble [†], U. Rajesh Kumar [‡,Δ], Raman Sankar [§,⊥], Zachary A. Bogorad [¶], Fang-Cheng Chou [§], Yit-Tsong Chen[‡,Δ] and Xuan P. A. Gao[\*,†]*

[†] Department of Physics, Case Western Reserve University, 2076 Adelbert Road, Cleveland OH 44106

[‡] Department of Chemistry and [§] Center for Condensed Matter Sciences, National Taiwan University, Taipei 10617, Taiwan

[⊥] Institute of Physics, Academia Sinica, Taipei 11529, Taiwan

[Δ] Institute of Atomic and Molecular Sciences, Academia Sinica, Taipei 10617, Taiwan

[¶] Solon High School, 33600 Inwood Dr, Solon, OH 44139

[\*] Email: (X.P.A.G.) xuan.gao@case.edu

**ABSTRACT** Graphene-like two-dimensional (2D) materials, not only are interesting for their exotic electronic structure and fundamental electronic transport or optical properties but also, hold promises for device miniaturization down to atomic thickness. As one material belonging to this category, InSe, a III-VI semiconductor, is not only a promising candidate for optoelectronic devices but also has potential for ultrathin field effect transistor (FET) with high mobility transport. In this work, various substrates such as PMMA, bare silicon oxide, passivated silicon oxide, and silicon nitride were used to fabricate multi-layer InSe FET devices. Through back gating and Hall measurement in four-probe configuration, the devices' field effect mobility and intrinsic Hall mobility were extracted at various temperatures to study the material's intrinsic transport behavior and the effect of dielectric substrate. The sample's field effect and Hall mobilities over the range of 20-300K fall in the range of 0.1-2.0×$10^3$ cm$^2$/Vs, which are comparable or better than the state of the art FETs made of widely studied 2D transition metal-dichalcogenides.

In the current stream of research in finding new materials to replace the conventional Si-based devices which is being widely used in logic circuits, researchers have explored a wide range of materials to overcome the scaling limit of the Si-based devices. Since the realization of device fabrication using single layer graphene and its 2D massless Dirac fermion,[1,2] there has been great interest in 2D graphene-like materials, in particular transition-metal dichalgenides (TMDs)[3-6] such as MoS$_2$,[7-11]

MoSe$_2$,[12-14] WS$_2$,[15] and WSe$_2$.[16,17] With the maximum carrier mobility of 2D TMD devices limited to a few hundred cm$^2$/Vs,[18-20] other 2D materials with similar layered 2D crystal structures are being sought after for better charge transport mobility and device performance.[21,22] A notable example is the recent demonstration of multi-layer black phosphorus (or phosphorene) FETs showing field effect mobility of holes approaching 10$^3$ cm$^2$/Vs.[21]

InSe is a 2D material made of stacked layers of Se-In-In-Se atoms with van der Waals (vdW) bonding between quadruple layers (Figure 1a). In the bulk form, InSe's mobility could be near 10$^3$ cm$^2$/Vs at room temperature ($T$) and exceeds 10$^4$ cm$^2$/Vs at low $T$,[23, 24] making it another promising candidate for the next generation high performance 2D semiconductor devices. Some recent works also highlight the potential applications of InSe and related III-VI 2D materials in optoelectronics. For example, Lei *et al.* and Tamalampudi *et al.* showed that devices of few-layer InSe obtained by mechanical exfoliation can be used as photo sensor with high photo-responsivity.[25, 26] Additionally, electroluminescence was observed in vertically stacked InSe/GaSe hetero-junction fabricated based on the mechanical exfoliation method for 2D vdW materials.[27] In terms of electrical transport device, it was recently studied by Feng *et al.* that with PMMA coated on Al$_2$O$_3$ as dielectric layer for InSe FET, the two-terminal room temperature field effect mobility of the sample can be improved to be ~1000 cm$^2$/Vs which is approaching its best Hall mobility value in the bulk[23] and well above that of the TMDs.[28] However, due to the limitation of two-terminal measurement and only room temperature behavior was studied, the intrinsic transport properties and mobility limiting mechanisms in such devices still remain to be understood.

In this letter, we report the electron transport characterization of multi-layer InSe devices and elucidate the effects of contact, temperature and different substrate dielectrics (SiO$_2$, Si$_3$N$_4$, Hexamethyldisilazane (HMDS) passivated SiO$_2$ and PMMA). By four-terminal measurement, the intrinsic transport properties and mobility were obtained. It is found that due to the inclusion of contact effect, the standard two-terminal FET configuration tends to underestimate the field effect mobility compared to the intrinsic value obtained by four-terminal measurements. Both the intrinsic field effect mobility and Hall mobility increase with decreasing temperature, indicating the relevance of phonon scattering in limiting the mobility of InSe nanoflakes. The dielectric property of substrate also plays a major role on the mobility of sample. While the PMMA dielectric substrate gives the best mobility (maximum Hall mobility ~2400 cm$^2$/Vs) among all the substrates, the typical value (100-2000 cm$^2$/Vs for $T$ between 20K and 300K) of mobilities obtained in multi-layer InSe compares favorably or better than TMDs.

Bulk InSe crystals were grown with the Bridgman method, similar to Ref.[26]. To grow high-quality single-crystalline InSe, we used 99.999% pure molar mixture of In and Se compound purchased from Sigma Aldrich. Synthesis of the crystals was carried out in a quartz ampoule by placing the mixture of the compounds at one end of the ampoule, evacuating the ampoule to ~$10^{-4}$ Pa, and subsequently sealing the other end of the ampoule. Homogenization of the mixture was conducted in a horizontal furnace at 600°C for 48 h. The as-grown crystals of excellent optical quality were easy to cleave to obtain crystalline planes perpendicular to the trigonal *c*-axis.

The multilayer InSe nanoflake samples were mechanically exfoliated onto degenerately doped silicon substrate with different dielectrics ($SiO_2$, $Si_3N_4$, PMMA and HMDS modified $SiO_2$) on surface using the standard scotch tape method. Four types of substrates were cleaned and prepared. For the substrates with bare $SiO_2$ and $Si_3N_4$, substrates were cleaned in boiling acetone at 100 C for 1 hour and then rinsed with ethanol and DI water following by blow drying with compressed air. Some Si/$SiO_2$ substrates were modified with HMDS (MicroChem™) to remove the water adsorbed on surface and render a charge neutral surface. For samples with PMMA dielectric layer on $SiO_2$, ~200 nm thick PMMA was spin-coated onto the Si/$SiO_2$ substrate followed by baking at 180°C for 30 minutes. It is note-worthy that the HMDS modified substrates had the lowest yield in exfoliation and the exfoliated InSe flakes are significantly smaller than other types of substrates, presumably due to the strong hydrophobicity of the surface.

For best mobility results, flakes that are roughly 20-40 nm thick were chosen for this study. To fabricate the electrodes contacting InSe nanoflake, a StrataTek™ copper grid was placed on top of the sample and used as a shadow mask for resist-free metal evaporation. Cr/Ag (10nm/80nm) was used as the contact metal in most devices and the typical distance between electrodes is 20 μm. Figure 1b and c illustrate the four-terminal and two-terminal device scheme, showing the InSe nanoflake, metal contact, dielectric layer and degenerately doped Si substrate which was also used as a back-gate to tune the carrier density. Optical images of devices are also shown. The prepared samples are loaded in a Lakeshore vacuum probe station and cooled down with liquid nitrogen to study the temperature dependence of the field effect mobility while some four-terminal samples were also loaded in Physical Property Measurement System (PPMS) for Hall measurements. Typical source-drain voltage (Vsd) used in the experiments was 0.1-1V.

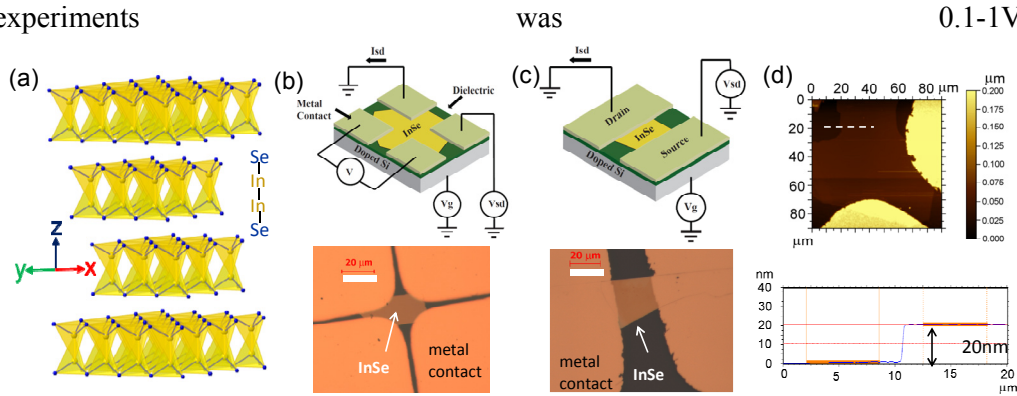

**Figure 1**. (a) Layered crystal structure of InSe (blue dots: Se atoms; yellow dots: In atoms). (b, c) Schematic (top) and optical image (bottom) of InSe nanoflake device

for backgating and mobility measurement in the four-terminal (b) and two-terminal (c) configuration. The scale bars in the optical images are 20 μm. (d) (top) AFM topography of a InSe nanoflake sample with 20 nm thickness (a line scan of height profile is shown in the bottom panel).

The samples' two-terminal current-voltage ($I_{sd}$-$V_{sd}$) characteristics were first characterized to check the quality of contacts at different temperatures and different strength of applied back gate voltage $V_g$. Figure 2a and b show typical $I_{sd}$-$V_{sd}$ curves of an as prepared sample at $T = 200$ K and 77 K. As can be seen, large positive gate voltage induces higher current, indicating n-type conduction in the InSe device. It is also evident that the $I_{sd}$-$V_{sd}$ curves at low $T$ are less linear (Ohmic) than high $T$ curves, showing the effect of Schottky barrier in limiting the current through metal-InSe interface at low $T$. The effect of Schottky barrier and non-ideal contact creates some differences between the conductance measured in standard two-terminal FET setup (Fig. 1c) and the intrinsic value. To illustrate, we show in Figure 2c and d a comparison of the two-terminal conductance and the four-terminal conductance for the same device (PMMA-2). With the influence of temperature and gate-voltage dependent contact resistance effects removed, the intrinsic four-terminal conductance is about ten times higher than the two-terminal value.

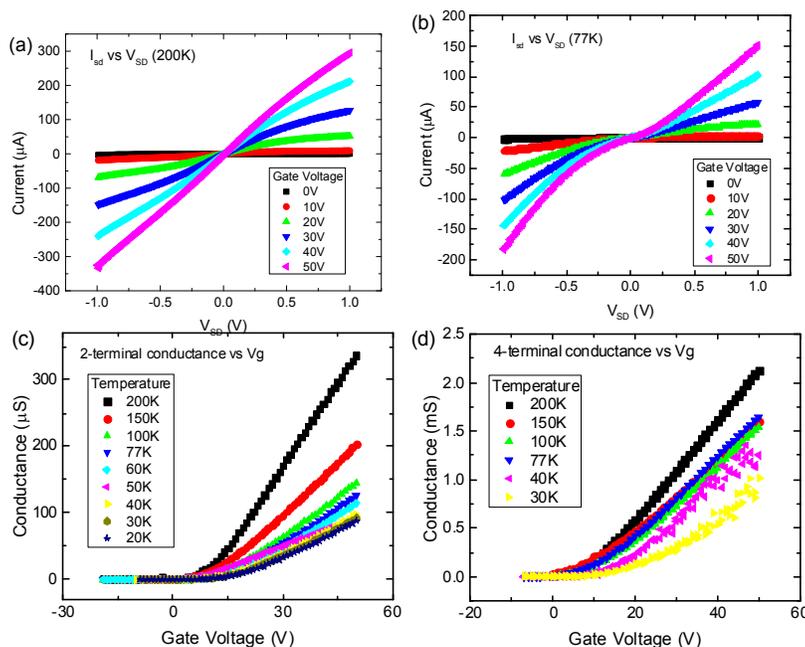

**Figure 2.** Typical I-V curves obtained from InSe device on PMMA covered Si/SiO$_2$ substrate (flake size ~ 81.9 x 76.5 μm$^2$) at 200 K (a) and 77 K (b). Variation of two-terminal conductance ($V_{sd} = 0.1$V) (c) and four-terminal conductance ($V_{sd} = 0.5$V) (d) as a function of back-gate voltage of InSe nanoflake device on PMMA covered Si/SiO$_2$ substrate at different temperatures.

The two-probe gate transfer curve is analyzed to extract several key metrics of FET. Our multi-layer InSe FETs have On-Off ratio on the order of $10^7$ and subthreshold swing (SS) on the order of 1V/decade (Supplementary Information Figure S1). The conductance $G$ vs. $V_g$ curves in Figure 2c and d can also be used to extract the field effect mobility $\mu_{FE}$ through the trans-conductance $g_m = dG/dV_g$. Related to the switching speed of FET, $\mu_{FE}$ is an important character of FET. For our planar few-layer InSe device, the two-terminal field effect mobility was calculated from the relation: $\mu_{FE} = \frac{L}{W} \frac{dI_{sd}}{C_i V_{sd} dV_g} = \frac{L}{W} \frac{g_m}{C_i}$, where $L$ and $W$ are the length and width of the sample, $C_i$ is the capacitance per unit area of the corresponding dielectric layer (which is given by $C_i = \frac{\epsilon_0 \epsilon_r}{d}$ where $\epsilon_0$ is free space permittivity, $\epsilon_r$ is the substrate's dielectric constant and $d$ is the dielectric layer's thickness). The extracted two-terminal and four-terminal peak $\mu_{FE}$ at different temperatures are included in Figure 3a, using calculated geometric $C_i$ listed in Table S1 of Supplementary Information. (The gate voltage dependence of field effect mobility can be found in Figure S2 of Supplementary Information). Likely due to the impact of worsened contacts at low $T$, the two-terminal field effect mobility continuously drops at lower $T$. The four-terminal field effect mobility shows a weak temperature dependence, in contrast to the $T$-dependent Hall mobility to be discussed later. This discrepancy is likely due to that the actual gate capacitance of PMMA-InSe device increases with $T$ between 200K and 300K, as shown by our Hall density measurement (see Figure S3 and related discussion in the Supplementary Information). Note that in calculating the four-terminal field effect mobility, we used the equation $\mu_{FE} = \frac{g_m}{C_i} / \frac{\pi}{ln2}$ to account for the square shaped geometry of the measurement, according to the relation between resistance per square and the directly measured resistance in the van der Pauw method.[29]

The high field effect mobility in FET device of InSe on PMMA at room temperature was attributed to the dielectric screening from PMMA in Ref. [28], similar to a previous work on $MoS_2$ FET.[11] Our experiments confirmed that InSe FET on PMMA substrate indeed has higher field effect mobility than other commonly used substrates. As shown in Figure 3b, typical $\mu_{FE}$ of two-terminal InSe FET on $SiO_2$, $Si_3N_4$ or HMDS modified $SiO_2$ substrates is 50-200 cm$^2$/Vs at room $T$ while InSe on PMMA has $\mu_{FE}$ higher than 1000 cm$^2$/Vs. The 2-terminal device with PMMA as dielectric layer shows mobility of 1250 cm$^2$/Vs at room temperature, well above any other type of substrates and generic values for TMDs. However, the InSe FET supported on PMMA also exhibits a rapidly decreasing trend in $\mu_{FE}$ as $T$ decreases while other conventional solid dielectrics yield increasing $\mu_{FE}$ as $T$ decreases. This may be related to be the more severe degradation of PMMA at cryogenic temperatures which caused worsened electrical contacts. Indeed, we more frequently experienced failure of PMMA supported InSe devices at low temperature than devices made on other substrates. Figure 3c plots the field effect mobility of a InSe device on $Si_3N_4$ dielectric for which we could perform both two-terminal and four-terminal measurements. Although the two-terminal $\mu_{FE}$ is lower than the four terminal value due to the contact effect, one sees that both mobilities show similar trend of increase with lowering $T$, a character of reduced phonon scattering in typical semiconductors. This suggests that conventional dielectrics do have the advantage of being stable against temperature cycling, despite the somewhat lower mobility compared with PMMA polymer. It is also worth noting that devices made directly on $SiO_2$ or $Si_3N_4$ substrates generally showed significant hysteresis in the conductance *vs.* gate voltage curve. The hysteresis effect reduces as the temperature is reduced (Supplementary Information, Figure S4). Such hysteresis is likely due to charge trapping on $SiO_2$/InSe interface or hydration on $SiO_2$ or nitride surface.[30-32] We observed weaker hysteresis in devices on PMMA or HMDS modified $SiO_2$, presumably owing to the surface being passivated and dehydrated.[33]

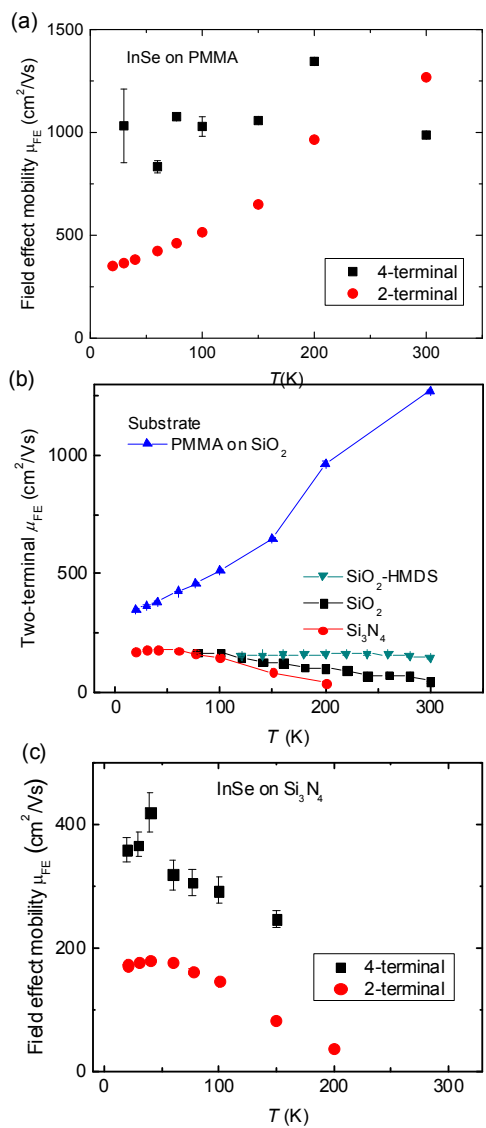

**Figure 3**. Comparison between the field effect mobility of two-terminal and four-terminal measurements of InSe on PMMA coated $Si/SiO_2$ substrate (a) and $Si/Si_3N_4$ substrate (c). (b) Comparison between the field effect mobility of two-terminal InSe FET devices on different substrates.

With the four contacts available in our square shaped four-terminal device, Hall effect measurement can be performed along with the resistance measurement to obtain the electron density $n$ and Hall mobility $\mu_H$. For Hall measurement, four-terminal devices were subjected to magnetic field $B$ applied in perpendicular direction to the sample surface where the transverse or Hall resistance $R_{xy}$ was extracted for Hall coefficient $R_H = R_{xy}/B$. In all devices measured, $R_H$ was determined from linear fitting $R_{xy}(B)$ data within ±2T magnetic field. The electron density was calculated using $n = -1/eR_H$, giving a range of 0.1-2.0x10$^{13}$ cm$^{-2}$ as presented in Figure 4a.

The Hall mobility was calculated using the equation $\mu_H = -\sigma R_H$, where $\sigma$ is the sheet conductance per square averaged over multiple directions as per the van der Pauw method. The Hall mobility *vs* temperature plot on samples with PMMA and Si$_3$N$_4$ dielectric is shown in Figure 4b. The sample on PMMA showed Hall mobility as high as 2000 cm$^2$/Vs below 100 K at $V_g$ = +50V ($n$~4×10$^{12}$/cm$^2$), much higher than the field effect mobility in either the two-terminal or four-terminal configuration (in Fig. 3). The Hall mobility data also revealed significant temperature dependence. Figure 4b shows an increasing trend of Hall mobility as the temperature decreases for both PMMA and Si$_3$N$_4$ substrates, reflecting the reduced phonon scattering effects as the temperature lowers. Meanwhile, similar to the gate dependence of the field effect mobility (Figure S2, Supplementary Information), the Hall mobility also varies significantly over the carrier density range studied (10$^{12}$-10$^{13}$/cm$^2$). From Figure 4, a sharply increasing Hall mobility of density is observed in PMMA supported InSe sample (e.g. $\mu_H$ at 50 K doubled from 1 to 2×10$^3$ cm$^2$/Vs as $n$ increased from ~2 to 4×10$^{12}$/cm$^2$). However, for silicon nitride supported device with density in the range of 10$^{13}$/cm$^2$, the Hall mobility change is weaker. Additionally, the fact that the Hall mobility increases sharply with the decreasing temperature also implies that the reduction of 2-terminal field effect mobility over the decreasing temperature shown in Figure 3a can be more likely due to the worsened contacts at lower temperature and the *T*-dependent gate capacitance rather than enhanced interface scattering between InSe and frozen PMMA. It is thus hopeful that such density and dielectric effects on the mobility could be further engineered to enhance the mobility.

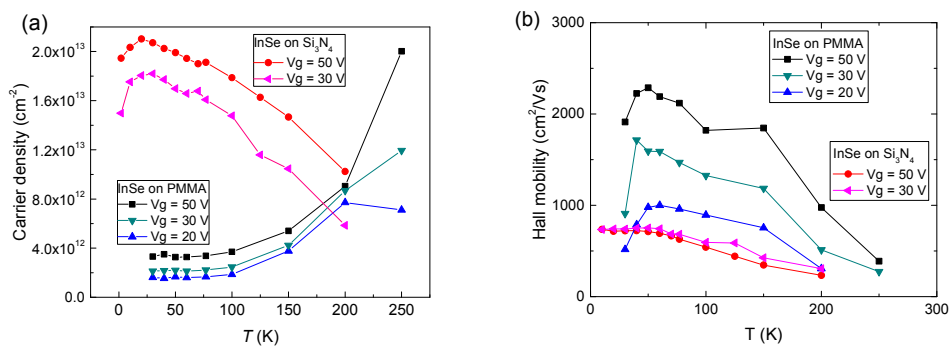

**Figure 4**. Carrier density (a) and Hall mobility (b) of InSe samples on PMMA and

$Si_3N_4$.

In summary, multi-layer InSe FET devices with two-terminal or four-terminal configuration were fabricated using shadow mask method on different kinds of dielectric materials. The effect of the dielectric on the samples' mobility was explored over a wide range of temperature. While PMMA substrate has the highest field effect mobility ($>10^3$ cm$^2$/Vs at room $T$), the intrinsic four-terminal field effect mobility of InSe falls in the range of 100-1000 cm$^2$/Vs at room temperature for all the dielectrics investigated (silicon oxide, nitride and HMDS passivated silicon oxide). The Hall mobility of InSe nanoflake also exceeds 2000 cm$^2$/Vs at low temperatures (~20K) and exhibits phonon scattering effect. Our work demonstrates the promises and potential of III-VI semiconductor InSe as a 2D material for high performance electronics.

ACKNOWLEDGMENT

X. P. A. G. acknowledges the NSF CAREER Award program (grant No. DMR-1151534) for financial support of research at CWRU. Y.T.C. acknowledges the Ministry of Science and Technology of Taiwan (grant No. MOST-103-2627-M-002-009) for financial support. F.C.C. acknowledges the support provided by the Ministry of Science and Technology of Taiwan (grant No. MOST-102-2119-M-002-004) and Academia Sinica under AC Nano Program_2014.

ASSOCIATED CONTENT

Supporting Information

Typical device's subthreshold swing, ON-OFF ratio, field effect mobility versus gate voltage, gate capacitance for different dielectrics, and effect of substrate dielectric on device's hysteresis (including Figure S1-S4 and Table S1). This material is available free of charge via the Internet at http://pubs.acs.org.

The authors declare no competing financial interest.

REFERENCES

1. Novoselov, K. S. *et al. Nature* **2005**, *438*, 197–200.
2. Zhang, Y.B., Tan, Y.W., Stormer, H.L., Kim, P. *Nature* **2005**, *438*, 201-204.
3. Wang, Q.H., Kalantar-Zadeh, K., Kis, A., Coleman, J.N., Strano, M.S. *Nature Nanotechnology*, **2012**, *7,* 699-712.
4. Butler, S.Z. *et al.*, *ACS Nano*, **2013**, *7*, 2898-2926.
5. Chhowalla, M., Shin, H.S., Eda, G., Li, L.J., Loh, K.P., Zhang, H., *Nature Chemistry* **2013**, *5*, 263-275.
6. Jariwala, D., Sangwan, V.K., Lauhon, L.J., Marks, T.J., Hersam, M.C. *ACS Nano*, **2014***, 8,* 1102–1120.
7. Radisavljevic, B., Radenovic, A., Brivio, J., Giacometti, V. & Kis, A. *Nature Nano.* **2011**, *6*, 147–150.

8. Yoon, Y., Ganapathi, K. & Salahuddin, S. *Nano Lett.* **2011**, *11,* 3768–3773.

9. Kim, S. *et al.*, *Nature Comm.* **2012**, *3*, 1011.

10. Das, S., Chen, H.Y., Penumatcha, A.V., Appenzeller, J. *Nano Lett.*, **2013**, *13* (1), 100–105.

11. Bao, W., Cai, X., Kim, D., Sridhara, K. & Fuhrer, M. S. *Appl. Phys. Lett.* **2013**, *102,* 042104.

12. Larentis, S., Fallahazad, B. & Tutuc, E. *Appl. Phys. Lett.* **2012**, *101,* 223104.

13. Chamlagain, B. *et al.*, *ACS Nano*, **2014**, *8* , 5079-5088.

14. Pradhan, N.P. *et al.*,   *ACS Nano*, **2014**, *8* , 7923–7929.

15. Sik Hwang, W. *et al.* , *Applied Physics Letters* **2012**, *101,* 013107.

16. Podzorov, V., Gershenson, M. E., Kloc, C., Zeis, R. & Bucher, E. *Applied Physics Letters* **2004**, *84,* 3301–3303.

17. Chuang, H.J, *et al.*, *Nano Lett.,* **2014**, *14 (6),* 3594–3601.

18. Fuhrer, M.S., Hone, J. *Nature Nano.,* **2013**, *8,* 146–147.

19. Radisavljevic, B, Kis, A. *Nature Mat.,* **2013**, *12,* 815–820.

20. Baugher, B.W., Churchill, H.O.H., Yang, Y., Jarillo-Herrero, P. *Nano Lett.* **2013**, *13*, 4212−4216.

21. Li, L. *et al.*, *Nature Nano.,* **2014**, *9,* 372–377.

22. Liu, H. *et al. ACS Nano* **2014**, *8,* 4033–4041.

23. Segura, A., Pomer, F., Cantarero, A., Krause, W. and Chevy, A. *Phys. Rev. B,* **1984***, 29*, 5708.

24. Savitskii, P.I., Kovalyuk, Z.D. and Mintyanskii, I.V. *phys. stat. sol. (a)* **2000***,* 180, 523-531.

25. Lei, S. *et al., ACS Nano* **2014**, *8*, 1263-1272.

26. Tamalampudi, S. R. *et al., Nano Lett.* **2014**, *14*, 2800–2806.

27. Balakrishnan, N. *et al., Adv. Opt. Mat.* **2014**, *2* (11), 1064-1069.

28. Feng, W., Zheng, W., Cao, W. & Hu, P., *Adv. Mater.* **2014**, *26*, 6587-6593.

29. van der Pauw, L. J. *Philips Research Reports*, **1958**, *13*, 1-9.

30. Kim, W. *et al*., *Nano Lett.* **2003**, *3*, 193–198.

31. Hang, Q. *et al.* , *Nano Lett.* **2008**, *8*, 49–55.

32. Wang, H., Wu, Y., Cong, C., Shang, J. & Yu, T. , *ACS Nano* **2010**, *4*, 7221–7228.

33. Joshi, P., Romero, H. E., Neal, A. T., Toutam, V. K. & Tadigadapa, S. A., *J. Phys.: Condens. Matter* **2010**, *22,* 334214.